# USING WORD CLOUDS FOR FAST IDENTIFICATION OF PAPERS' SUBJECT DOMAIN AND REVIEWERS' COMPETENCES[15]

**Yordan Kalmukov, PhD**
Department of Computer Systems and Technologies,
University of Ruse
E-mail: jkalmukov@ecs.uni-ruse.bg

***Abstract:*** *Generating word (tag) clouds is a powerful data visualization technique that allows people to get easily acquainted with the content of a large collection of textual documents and identify their subject domains for a matter of seconds, without reading them at all.*

*This paper suggests applying word clouds visualization to conference management systems (specialized document management and decision support systems) in order to support and facilitate decision making in at least two important processes – forming the Programme Committee by inviting suitable reviewers and manual (re)assignment of reviewers to papers. Word clouds proved to be very useful tool for fast identification of papers' subject domain and reviewers' competences.*

## INTRODUCTION

Generating word (tag) clouds is a powerful data visualization technique that allows people to get easily acquainted with the content of a large collection of textual documents and identify their subject domains for a matter of seconds. Instead of reading many files, their content could be automatically extracted and processed, so that only the most significant terms are retained and presented in a form of a word cloud. Usually the most significant and frequently appearing words are visualized in different colors and higher font sizes so they catch attention immediately (fig. 1).

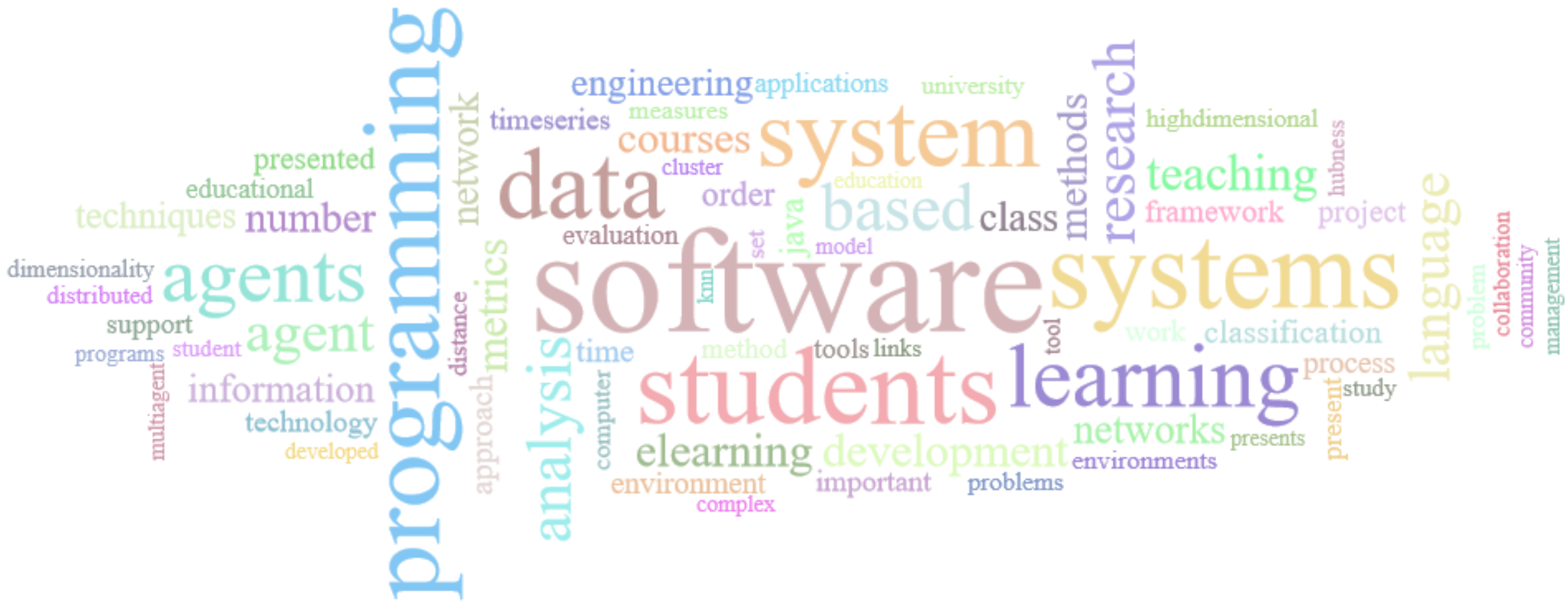

Fig. 1. Word cloud showing the areas of research of more than 100 scientific papers

Word clouds have been used for years in various subject domains, most commonly in web-based information systems such as content management systems, crowdsourcing platforms, social networks and others.

[15] Докладът е представен на заседание на секция 3.2 на 29 октомври 2021 с оригинално заглавие USING WORD CLOUDS FOR FAST IDENTIFICATION OF PAPERS' SUBJECT DOMAIN AND REVIEWERS' COMPETENCES

However, this data visualization method could be successfully applied in conference management systems (CMS) or in journal editorial managements systems as well, to support and facilitate at least two important processes:

- Invitation of suitable reviewers to join the Program Committee (PC)
- Manual assignment and reassignment of reviewers to papers

These processes are not supposed to be automated, but done my human experts – the PC chairs. However, any automatically gathered, processed and aggregated information that could support decision making is welcome. Word clouds could be generated from the titles and the abstracts of submitted papers and the papers, previously published by the reviewers. In this sense, word clouds could promptly reveal to PC chairs the thematic fields of all submitted papers or the research interests / competences of the PC members (reviewers) who accept to evaluate them.

## EXPOSITION

### Related work

It could not be easily determined how much, or whether at all, word clouds are used in conference management systems. I do not know any commercially available or experimental conference management system applying this data visualization technique. Bertolino et al. (2018) use word clouds generated from the titles of the papers published at ASE, FSE, and ICSE conferences in order to identify the topics of highest interests to the authors, i.e. topics with the highest number of submitted papers.

In contrast, word clouds are frequently used in social networks and crowdsource platforms.

Wang et al. (2020) propose semantics-based word cloud visualization to be applied on user reviews of products and services. Reviews by other users are very important for decision making where to buy one product or another. However when their number gets large, it would be very time-consuming for people to read them all. So they could be summarized in word clouds. Authors propose not to use ordinary random-layout clouds, but to apply a grammatical dependency parsing to identify semantically related words and draw them closer together. This indeed enhances word clouds and makes them more easily interpretable.

H. Jung (2014) also suggests using word clouds for easily interpretation of user reviews. Authors propose applying collapsed dependencies and sentiment analysis to build an interactive word cloud that can show grammatical relationships among words and display positive and negative polarity of each word.

Saranya and Geetha (2020) propose generating the word cloud from the topics derived by the LDA topic modelling algorithm. They apply their method on user reviews related to clothes.

Seifert et al. (2011) propose applying word clouds for easier document labelling for training data. Word clouds could be used as condensed representations of text documents, instead of the full-text document, and thus to reduce the labelling time needed for human labellers.

Giannoulakis and Tsapatsoulis (2021) use word clouds for topic visualization, where topics were derived from social media (Instagram) hashtags used for image annotation.

### Composing the Programme Committee by inviting suitable reviewers in the necessary areas of research only

Multidisciplinary conferences usually maintain similar Program Committee (PC) during the years. Of course it evolves in time and is being extended but not rapidly. The PC contains reviewers (almost) evenly distributed across all topics covered by the conference. However, the distribution of submitted papers in thematic fields and conference topics could not be predicted at all. It is possible to have plenty of submitted papers in one topic and just a few papers in another. So the number of experts in the first topic should be many more than the number of experts in the second one. Otherwise the accuracy of the assignment may get significantly lower and many papers may be evaluated by reviewers who are not experts in the relevant subject domains. If this happens, it could have quite a negative effect on the conference’s public image and the authors thrust in it. To prevent that, PC chairs could generate a word cloud, built from the abstracts of all submitted papers (fig. 2 and fig. 3) *in order to identify their subject domains*, and then *invite reviewers in those subject domains with higher priority*.

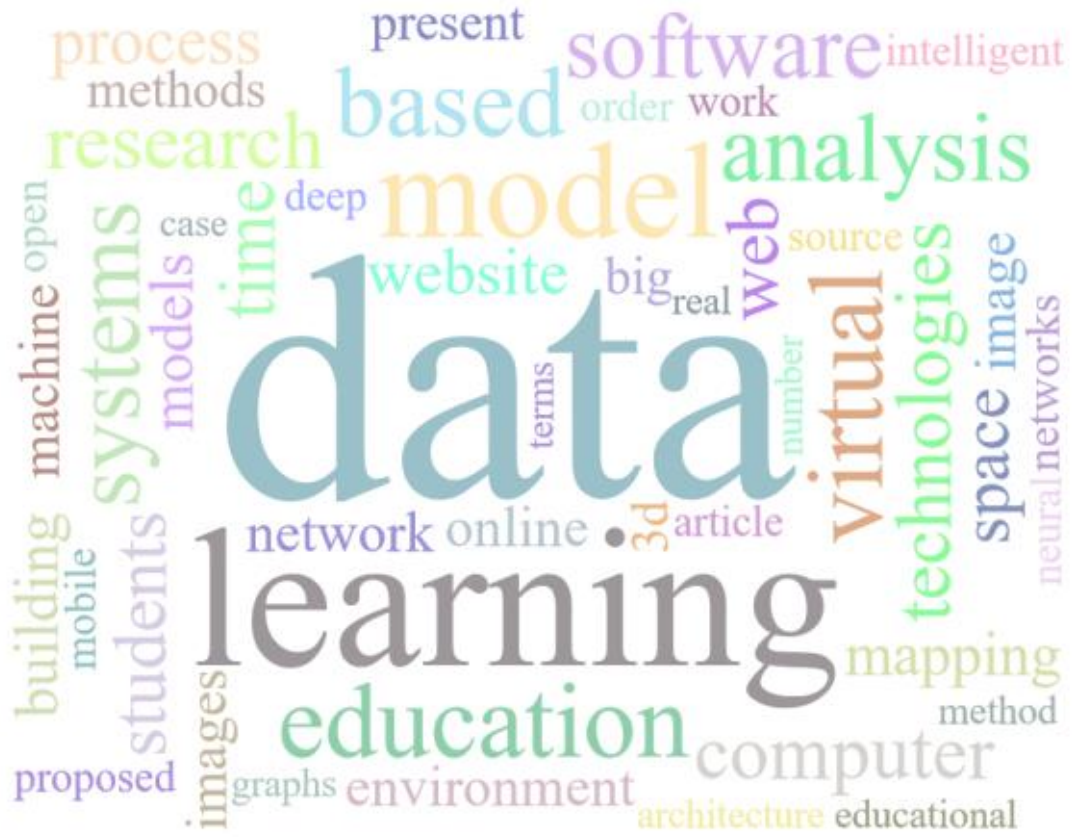

Fig. 2. Word cloud generated by the abstracts of 75 papers, submitted to CompSysTech 2018 conference

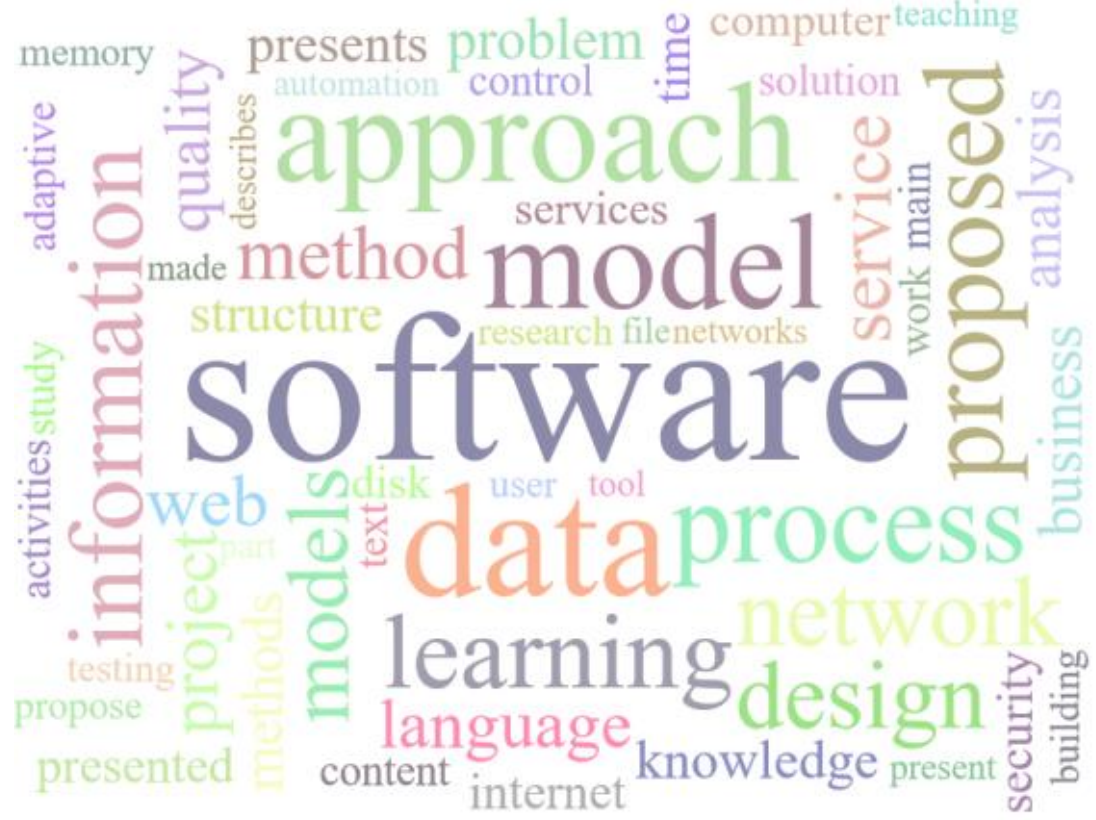

Fig. 3. Word cloud generated by the abstracts of 95 papers, submitted to CompSysTech 2012 conference

Alternatively, the word cloud could be built by the “conference topics”, selected by the authors during the paper submission. Many conferences, including CompSysTech ([www.compsystech.org](www.compsystech.org)), rely on explicit method for describing papers and reviewers’ competences by choosing keywords (usually conference topics) from a predefined list or taxonomy of topics. Although these topics are just single words, rather than larger amount of text, they could be also reliable source of identification information. Figures 4 and 5 show example of clouds, generated by the *conference topics* used to describe submitted papers in CompSysTech 2018 and 2012 respectively.

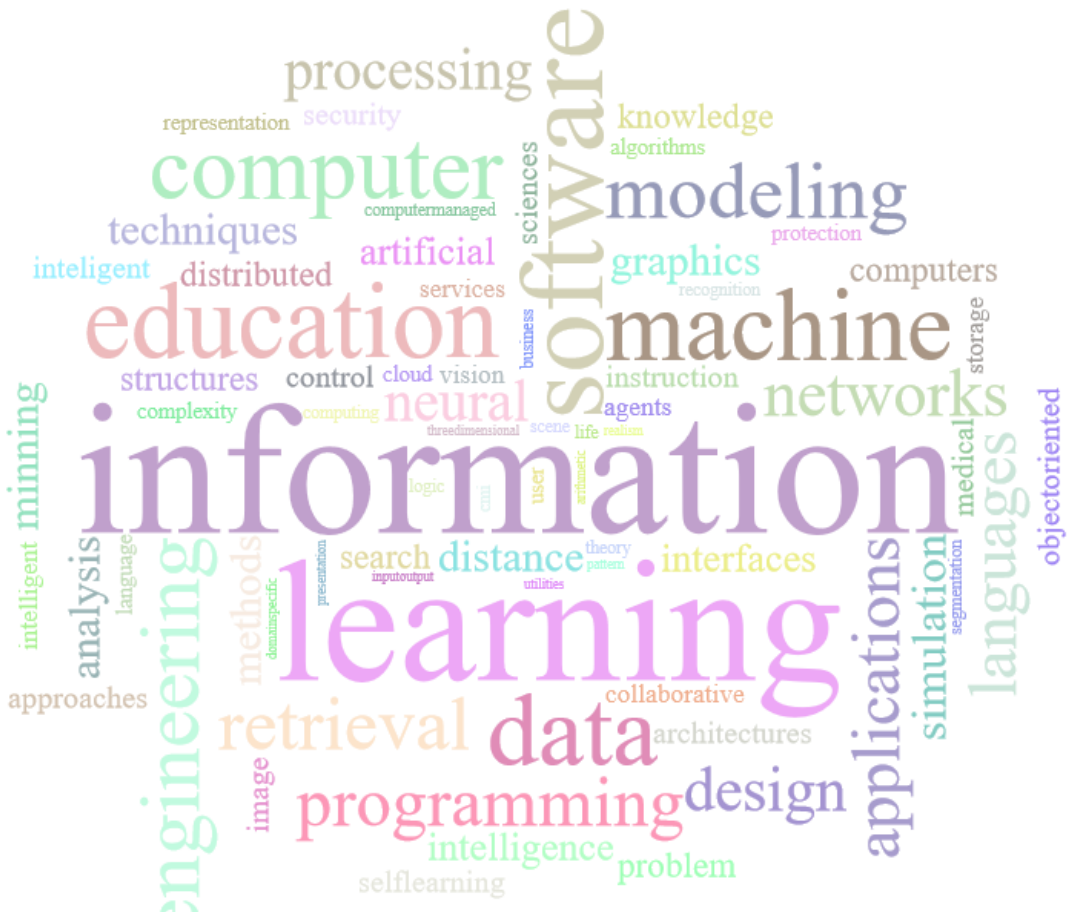

Fig. 4. Word cloud generated by the conference topics, describing 75 papers, submitted to CompSysTech 2018 conference

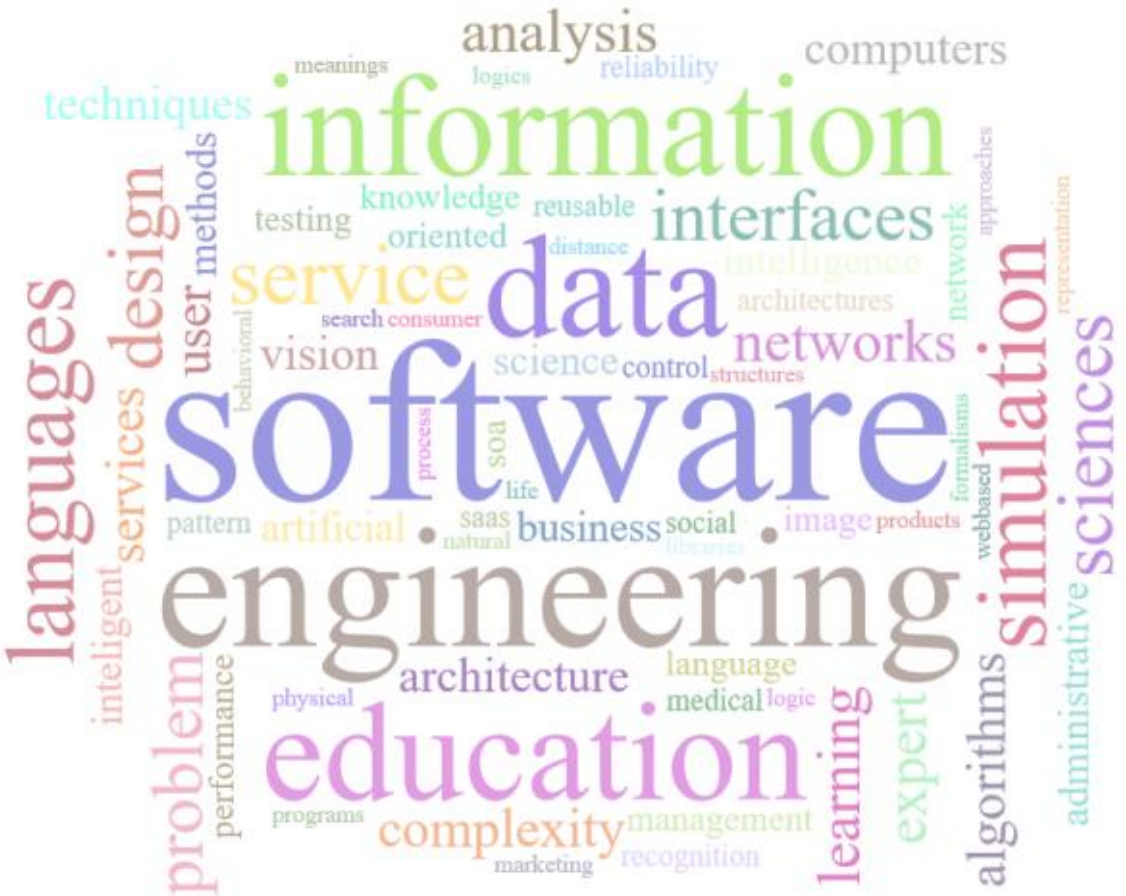

Fig. 5. Word cloud generated by the conference topics, describing 95 papers, submitted to CompSysTech 2012 conference

As seen, there is a tight correlation between figures 2 and 4, as well as figures 3 and 5. It shows that both, the abstracts and the conference topics, could be used for identification of the subject domains having the highest number of submitted papers. Furthermore, the high level of correlation also indicates that authors have actually selected the most suitable keywords (conference topics) to describe their papers. And also that the keywords (conference topics), composing the predefined list, are properly chosen – their number is not too few to lose focus, but also not too high to prevent clustering.

**Manual assignment and reassignment of reviewers**

In most cases the assignment of reviewers to papers is done *automatically* as the manual assignment, especially in case of large-scale multidisciplinary conferences, leads to quite inaccurate results due to the many constraints (accuracy, load balancing, conflict of interests and etc.) that should be taken into an account. However, there are usually papers that need be later reassigned manually to other reviewers. Why? Mostly because of:

- *Unreliable reviewers*, who do not complete their reviews on time, despite their commitment and promise to do so.
- *Need for arbitration* due to highly discrepant reviews.
- Hidden *conflict of interests* of the author(s) with a reviewer.

If the assignment algorithm supports incremental assignment, then it could be used for automatic reassignment of the papers to additional reviewers. Heuristic algorithms usually do, while the Hungarian algorithm of Kuhn and Munkres does not (Kalmukov and Rachev, 2010). But even if the algorithm supports it, the automatic reassignment is not the best option. Why? Because reassignment means an additional workload for some reviewers, while not everybody will accept it just a couple of days before the review submission deadline. So, it is better reassignment to be done manually since PC chairs know reviewers and know who are reliable, who will accept additional workload and will complete the reviews on time. They do not have time for trials and errors. However, to perform an accurate manual reassignment, the PC chairs should know the areas of expertise of all reviewers, which is actually not possible. To facilitate the process and increase its accuracy, PC chairs could generate a word cloud for each reviewer, built from the abstracts of his/her publications available on the Internet (figures 6 and 7). This is a fast and easy way to identify the areas of research and expertise of a reviewer.

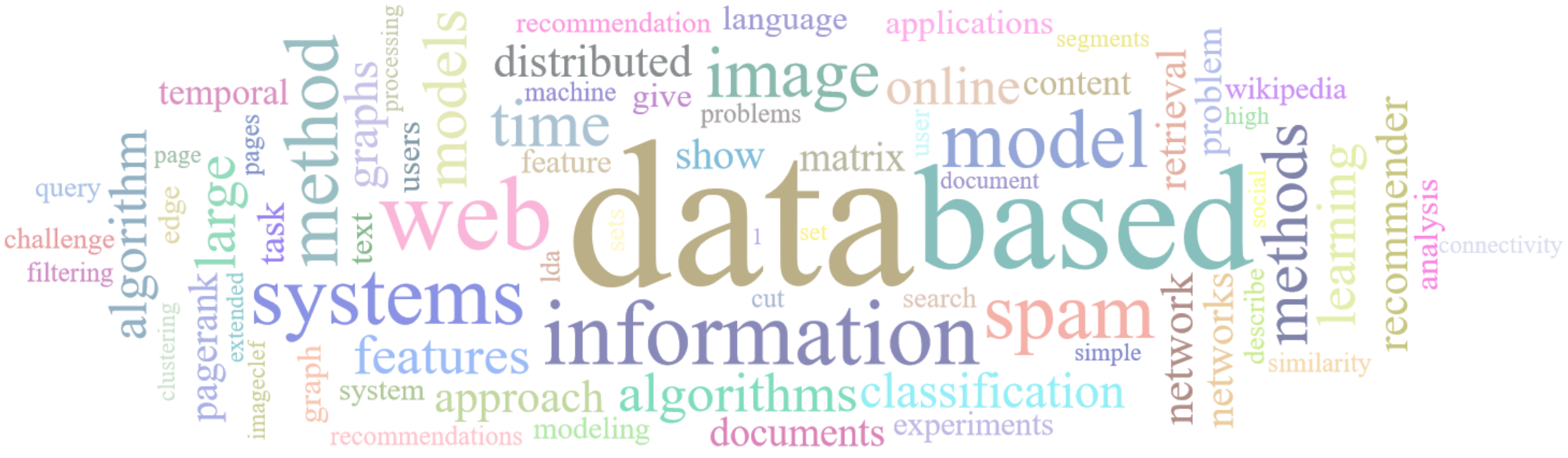

Fig. 6. Word cloud generated by the abstracts of **117** scientific papers of a CompSysTech reviewer

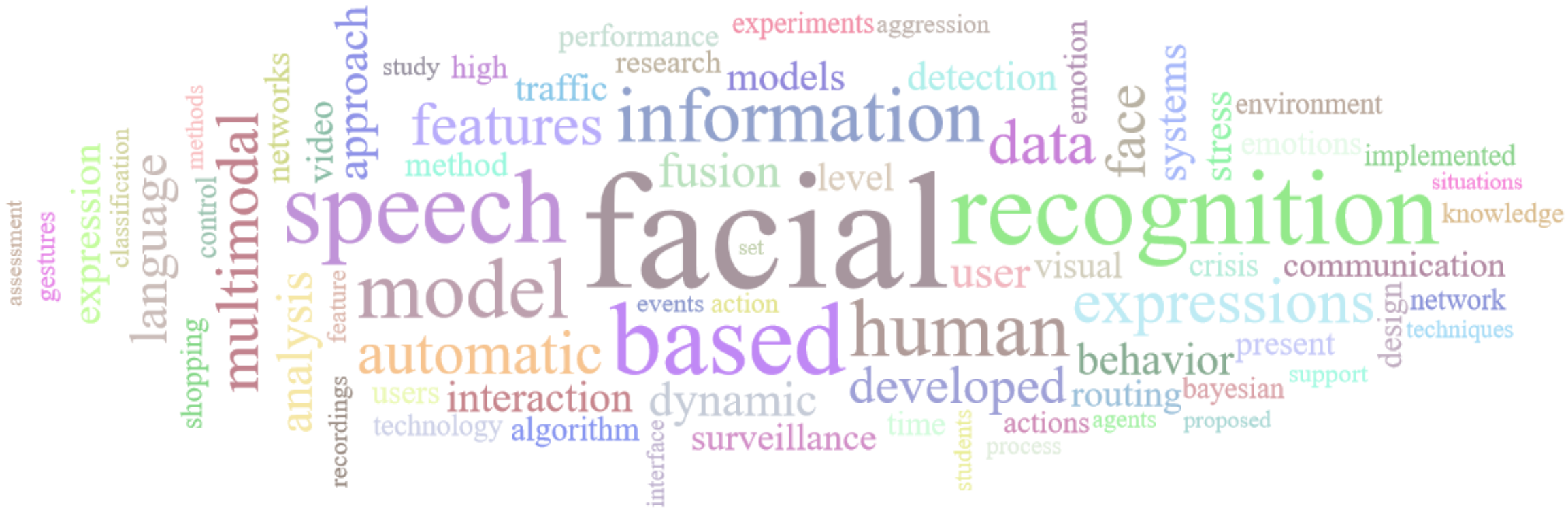

Fig. 7. Word cloud generated by the abstracts of **158** scientific papers of another reviewer

The real names of the reviewers are not mentioned in fig. 6 and 7 captions in order to comply with the latest data protection policies, although scientific interests are public data as well as the CompSysTech PC members' names.

To build word clouds for the reviewers, the software needs to fetch the abstracts of their previously published manuscripts. Fortunately all abstracts are freely available on the Internet and could be automatically obtained by publication indexing services such as Semantic Scholar (www.semanticscholar.org), DBLP (dblp.org) and others.

### Data processing and word cloud generation

Since word clouds are generated from free-style text (unstructured data), it should be pre-processed and “sanitized” first. The text will surely contain punctuations and semantically insignificant words such as prepositions, conjunctions, adverbs, pronouns and others. Yes, they are very important from a syntactic point of view, but they do not reveal any information about the text semantic and its subject domain. Instead, they just cause troubles - since they appear frequently, they notably lower the weight of other, semantically significant, words. So, all punctuation marks and semantically-insignificant parts of speech should be removed. These words are usually provided in the form of “stopwords” list.

Generating word clouds could be implemented in various programming languages. There are two important activities – text processing and drawing the cloud itself. Figures 1 to 7 are generated by the conference management system (Kalmukov, 2011) which I develop and support. It is implemented in php and the cloud is drawn by the JavaScript library awesomeCloud v0.2 (jQuery plugin) written by Russ Porosky.

## CONCLUSION

Word clouds represent a powerful data visualization technique that allows people to get easily acquainted with the content of a large collection of textual documents and identify their subject domains for a matter of seconds, without reading them at all. It proved to be very helpful for Programme Committee chairs when implemented in conference management systems. Word clouds enable them to quickly and easily identify:

- The subject domains (conference topics) with the highest number of submitted papers.
- The areas of research and competences of the individual reviewers.

This aggregated and timely presented information supports decision making and facilitates the relevant processes of filling the Programme Committee with suitable (for the submitted papers) reviewers and manual (re)assignment of reviewers to papers.

Fast identification of subject domains and areas of research is important for other processes in conference management as well. Thus, word cloud proved to be valuable asset in conference management in general.

## ACKNOWLEDGEMENT

This paper is supported by project 21-FEEA-01 “Intelligent computer systems: research of their evolution, application, and management”, funded by the Research Fund of the “Angel Kanchev” University of Ruse.